\documentclass{aa}
\usepackage{graphicx}
\usepackage{natbib}
\bibpunct{(}{)}{;}{a}{}{,} % to follow the A&A style
\begin{document}
   \title{Comparing various multi-component global heliosphere models}

   \authorrunning{H.-R. M\"uller et al.}
   \titlerunning{Comparing multi-component global heliosphere models}

   \author{H.-R. M\"uller\inst{1,2}%\fnmsep\thanks{also at: IGPP-UCR, University of California, Riverside}
          \and
          V. Florinski\inst{2}
          \and
          J. Heerikhuisen\inst{2}
          \and
          V.V. Izmodenov\inst{3,5}
%          V.V. Izmodenov\inst{3}\fnmsep\thanks{also at: IKI, Russian Academy of Sciences, and Institute of Problems in Mechanics, Russian Academy of Sciences}
          \and
          K. Scherer\inst{4}
          \and
          D. Alexashov\inst{5}
          \and
          H.-J. Fahr\inst{6}          }

   \offprints{H. M\"uller}

   \institute{Department of Physics and Astronomy, Dartmouth College,
              Hanover NH 03755, USA\\
              \email{hans.mueller@dartmouth.edu}
         \and
              Institute of Geophysics and Planetary Physics, University of California, Riverside CA
              92521, USA\\
              \email{vflorins@ucr.edu; jacobh@ucr.edu}
         \and
              Lomonosow Moscow State University, Department of Mechanics and Mathematics \& Institute of Mechanics,
              Moscow 119899, Russia\\
              \email{izmod@ipmnet.ru}
         \and
             Ruhr-Universit\"at Bochum, Institut f\"ur Theoretische Physik IV:
             Weltraum- und Astrophysik, 44780 Bochum, Germany\\
             \email{kls@tp4.rub.de}
         \and
             Institute for Problems in Mechanics, Russian Academy of Sciences,
              Moscow 119526, Russia%\\
         \and
             Argelander Institute for Astronomy, Dept. of Astrophysics, University of Bonn, Auf dem H\"ugel
             71, 53121 Bonn, Germany\\
             \email{hfahr@astro.uni-bonn.de}
                          }

  \date{Received September 19, 2007; accepted March 31, 2008}

  \abstract
{Modeling of the global heliosphere seeks to investigate the
interaction of the solar wind with the partially ionized local
interstellar medium. Models that treat neutral hydrogen
self-consistently and in great detail, together with the plasma,
but that neglect magnetic fields, constitute a sub-category within
global heliospheric models.}
{There are several different modeling strategies used for this
sub-category in the literature. Differences and commonalities in
the modeling results from different strategies are pointed out.}
{Plasma-only models and fully self-consistent models from four
research groups, for which the neutral species is modeled with
either one, three, or four fluids, or else kinetically, are run
with the same boundary parameters and equations. They are compared
to each other with respect to the locations of key heliospheric
boundary locations and with respect to the neutral hydrogen
content throughout the heliosphere.}
{In many respects, the models' predictions are similar. In
particular, the locations of the termination shock agree to within
7\% in the nose direction and to within 14\% in the downwind
direction. The nose locations of the heliopause agree to within
5\%. The filtration of neutral hydrogen from the interstellar
medium into the inner heliosphere, however, is model dependent, as
are other neutral results including the hydrogen wall. These
differences are closely linked to the strength of the interstellar
bow shock. The comparison also underlines that it is critical to
include neutral hydrogen into global heliospheric models.}
   {}


   \maketitle
%
%________________________________________________________________

\section{Introduction}

The interstellar medium in the immediate solar neighborhood is
part of the ``local interstellar cloud'' (LIC). The flow of the
partially ionized LIC past the Sun constitutes a pressure that
balances and terminates the expansion of the coronal solar wind.
These two winds create a morphology that includes the termination
shock transition of the supersonic solar wind to a hot heliosheath
or heliotail flow. An interstellar bow shock is likely to be
necessary as well to decelerate the LIC flow. The ionized flows of
the LIC and the solar wind are separated by the heliopause. The
LIC also supplies the system with interstellar neutrals,
predominantly with neutral hydrogen (H). Neutral H interacts
weakly with the plasma, mainly through charge exchanges with
plasma protons.

The distance even to the termination shock is large enough that
there are only a few in-situ measurements to date in the outer
heliosphere. Notable sources of information are the two Voyager
spacecraft at a distance of 104 AU and 84 AU from the Sun (2007
September), respectively, with Voyager 1 having passed into the
heliosheath region on 2004 December 16 \citep[e.g., ][]{Stone05},
and Voyager 2 on 2007 August 30. For examples of in-depth analyses
of observations relating to the outer heliosphere, we refer to
other contributions in this special issue
\citep{Bzowski08,Pryor08,Richardson08,Slavin08}.

Data from the outer heliosphere are sparse, and numerical modeling
of the global heliosphere/LIC system plays an important role for
the analysis and interpretation of observations. It is needed to
relate the undisturbed LIC flow and its physical parameters to the
processed and changed flow that we observe in the heliosphere
inside the termination shock. In fundamental ways all the LIC
constraints formulated in the accompanying papers
\citep{Bzowski08,Pryor08,Richardson08,Slavin08} involve global
heliosphere modeling. Also the evaluation of future data sets from
the Interstellar Boundary Explorer (IBEX) mission, which focuses
on secondary neutrals created in the heliosphere and on the LIC
oxygen and helium flow through the heliosphere \citep{IBEX},
depends crucially on this kind of modeling.

All such global models make assumptions and simplifications, most
often with the goal of isolating the influence of a specific
physical effect (e.g., the tilt of the LIC magnetic field with
respect to the LIC flow vector), or in order to keep computation
times reasonable. The identification of heliospheric asymmetries
with respect to the helium LIC flow vector \citep[][and references
therein]{Moebius04,Lallement05} has increased interest in the
development of realistic, three-dimensional (3D) MHD models, as
different orientations and strengths of the interstellar magnetic
field can help to explain these asymmetries. However, the fact
remains that neutral interstellar H entering the heliosphere has a
more decisive influence on the heliospheric shape, extent, and
particle content. For this reason, we focus here on numerical
models that treat the plasma/neutral interaction in a
self-consistent way, but neglect the influence of interplanetary
or interstellar magnetic fields. The models are, in principle, 3D
plasma/neutral codes for which plasma and neutrals are coupled by
charge exchange. Wherever it is possible, the assumption of
azimuthal symmetry reduces the numerical methods effectively to 2D
while still calculating the 3D heliosphere. The results of our
investigation will be also applicable to 3D MHD models (for a
recent overview, see \citet{Pogorelov08}), as long as the latter
also include neutrals self-consistently, as is essential for
models of the global heliosphere.

The charge exchange interaction is weak enough that the mean free
path lengths of neutral H are often large compared to typical
heliospheric distances (see the discussion in section 3.2).
Neutral H is thought to be in local thermodynamic equilibrium in
the LIC, but as charge exchange proceeds in the heliosphere,
secondary neutrals arise, and they also can exchange charge with
plasma ions. This effectively drives neutrals out of equilibrium
in the heliosphere, and plasma and neutrals equilibrate again only
far away from the heliosphere.

In H-p charge exchange, the newly born (secondary) neutral H has
the velocity characteristics of the plasma protons at the location
of interchange. However, the neutral is no longer bound to the
plasma flow and follows a simpler trajectory than the underlying
plasma parcel.  Due to this, it is convenient to sort the neutrals
into different populations depending on their origin. We will
label here as component 1 the primary neutrals directly from the
ISM as well as those born in charge exchange outside the bow
shock. Component 2 are those secondary neutrals that are created
by charge exchange between bow shock and heliopause. They reflect
the conditions of the warmer interstellar plasma decelerated in
the bow shock. Because of the deceleration, there is a neutral
density increase downstream of the bow shock, the hydrogen wall.

We label as component 3 neutrals those that are born from the hot
heliosheath and heliotail plasma, and component 4 those born in
the supersonic solar wind between the Sun and the termination
shock. Component 3 neutral velocities are dominated by the large
thermal proton velocity of the heliosheath and heliotail, and
hence their direction is mostly random. Since a fraction of
component 3 neutrals are directed to the innermost heliosphere and
can be detected as energetic neutral H, this whole component 3 is
often referred to as ``heliospheric ENA'' (energetic neutral
atoms). The fourth neutral component has recently been called
``neutral solar wind'' (NSW) because its cold, fast velocity
characteristics are similar to those of the supersonic wind of the
inner heliosphere. Note that in spite of its name, the NSW as
defined here is distinct from the neutral hydrogen originating
from the Sun \citep[e.g.,][]{BlumFahr,Olsen94}, which has been
called neutral solar wind earlier as well.

In the non-MHD models that are currently applied to the global
heliosphere problem there is agreement that due to the
out-of-equilibrium nature of neutral H it needs to be modeled
separately from the ionized matter. The plasma is commonly modeled
by gas-dynamic methods. There are two different popular methods
for treating the neutrals, to be coupled to the plasma in a
self-consistent way. The first method is kinetic, where particle
methods such as Monte-Carlo simulate the neutral populations on a
Boltzmann-microscopic level. The kinetic treatment is motivated by
the usually large mean free paths of neutrals. The second
approach, the multi-fluid method, is to simulate each of the four
neutral components as a separate fluid on an Euler-macroscopic
level, and assumes that the superposition of the resulting four
Maxwellian distributions represents the true, generalized
distribution function of heliospheric neutral H well. Sometimes,
fluid models are being restricted further (by choice) by
decreasing the number of fluids to less than four, as in the
\citet{Zank96} multi-fluid model (component 1 and 2 combined into
one fluid) and the \citet{Fahr00} Bonn model (components 1--4
combined, but fluids describing pickup ions and cosmic rays
introduced).

Without going into any detail, the two neutral modeling approaches
can be summarized as follows. The main advantage of the kinetic
approach is that it does not restrict the shape of the neutral
distribution function, and thus allows the irregularity of the
neutral distribution in the heliosphere to persist everywhere in
the heliosphere. The main advantage of the fluid approaches is
that they are orders of magnitude faster computationally, and that
their usual field variables (density, velocity, pressure) are
smooth down to the grid and timestep resolution. The main
disadvantage of particle kinetic methods is that their accuracy is
driven by particle statistics, i.e.\ to increase the accuracy of
results for a particular location at a particular time, more
particles have to be generated to coincide there at the desired
time. Both kinetic models below (section 3) employ variable
particle weights to allow trajectories to be split, leading to a
significant improvement in the statistical accuracy (to a targeted
$\sim$2\% level) at reasonable computational costs. The splitting
procedure used in the \citet{Baranov93} model is described by
\citet{Malama91}, while the \citet{Heerikhuisen06} model uses a
similar method based on splitting during charge exchange. The main
disadvantage of fluid models is that each neutral component is
forced into local thermal equilibrium, which, at the very least,
constitutes a loss of information (Maxwellians instead of a more
general distribution).

Two studies have recently engaged in detailed comparisons of
kinetic models versus multi-fluid gasdynamic models of neutral
hydrogen in the heliosphere, and put forward some of the possible
physical reasons for the differences that invariably occur
\citep{Alexashov05,Heerikhuisen06}.
For their study comparing global heliospheric models,
\citet[][hereinafter AI05]{Alexashov05} set out using the Moscow
kinetic code developed over the years in Moscow starting from the
original \citet{Baranov91} kinetic-gasdynamic model.
%In it, the plasma is modeled gas-dynamically, and the neutral
%hydrogen atoms are followed kinetically.
A certain set of solar wind and interstellar boundary parameters
is used throughout their study. AI05 compare the kinetic result,
and non-self consistent variants of it, to multi-fluid models in
which neutral H is modeled by one to four fluids, coupled
self-consistently to the same gas-dynamic plasma code used also
for the kinetic model. They find that kinetic and multi-fluid
models never agree completely. The agreement with the kinetic
method is best for the four-fluid model, and worst for the
one-fluid model. The boundary locations are further out in the
four-fluid case when compared to the kinetic model, namely, the
upwind bow shock (BS) by 4\%, the termination shock (TS) by 5\%,
and the heliopause (HP) by 9\%. The hydrogen wall material is more
decelerated and consequently the peak density is larger, and the
filtration more severe (less H entering through the termination
shock). This discrepancy is started by the kinetic model having a
weaker (plasma) bow shock than the four-fluid model, which in turn
is likely caused by more secondary neutrals passing to the
interstellar side of the bow shock than in the fluid case, as
displayed by AI05.

In a similar investigation, \citet[][hereinafter
HFZ06]{Heerikhuisen06} use their own kinetic code and compare the
result with their own version of a four-fluid model. They, too,
find a weaker bow shock in the kinetic case compared with the
fluid case, and the same chain of consequences, including a larger
H density in the hydrogen wall and a smaller density passing
through the termination shock for the fluid case. While TS and HP
are farther out for the fluid case, the BS is less far than in
their kinetic model. At four representative locations in the
heliosphere, HFZ06 compare the parallel velocity distribution
function between kinetic and multi-fluid models, the latter being
a superposition of 4 individual Maxwellians. At least for those 4
points on the stagnation axis, they find that the two interstellar
neutral components coincide very well with the Maxwellians from a
four-fluid model, and only the two heliospheric neutral components
deviate. AI05 and HFZ06 agree that the NSW component is much
hotter (i.e., broader velocity distributions) in the kinetic model
than in the fluid model. The ENA-``fluid'' is the most problematic
to be fit by a Maxwellian, at least outside the inner heliosheath
and the heliotail.

HFZ06 also compare their results to those obtained by AI05 with
their codes. This comparison is possible because HFZ06 use the
same boundary parameters. The two four-fluid codes correspond very
well to each other, minus a subtle difference that comes about by
the different internal treatment of the bow shock in the two
plasma codes. The two kinetic code results also differ in hydrogen
density between bow shock and heliopause, which again might have
to do with the internal treatment of the bow shock in the
underlying plasma codes.

In this paper, we take one step further back and compare the
results of sophisticated, comparable global heliosphere models
(albeit all axisymmetric and non-MHD), run on the same boundary
data set characterizing the solar wind at 1 AU and the pristine
interstellar medium. Since the modeling strategies even for the
plasma gas-dynamic model are different across the four groups
considered (they are compared in section 2), the differences
between the models are going to be larger than for the case of the
internal comparisons of AI05 and HFZ06. In this sense, the paper
focuses not on discovering additional physical reasons for
differences between the kinetic and the multi-fluid approach.
Rather, we are trying to state quantitatively how far apart or
close some key results are, in order to give the wider community a
sense how accurate statements derived from current neutral/plasma
models likely are.

The models used within the observational contributions of this
special issue all are related to the global models outlined below,
and all make use of specific additions or modifications depending
on the specific issues addressed.  The Richardson and Wang
one-dimensional MHD model \citep{Richardson08} concentrates on the
solar wind - interstellar flow interaction to describe the
slowdown in the supersonic solar wind. It does not include many of
the intricacies of the global models beyond the termination shock
as discussed below, but incorporates the detailed solar wind
temporal structure to come to a meaningful comparison between
inner and outer heliosphere. \citet{Bzowski08} take the
interstellar flow from a kinetic model similar to the one in AI05,
and then add a Monte Carlo calculation of the history of
individual particle trajectories in the inner heliosphere to get
the pickup ion characteristics between 1 and 5 AU. For
\citet{Pryor08} it is important to add the radiation transport of
solar Ly-$\alpha$, including multiple scattering on a global
heliospheric model. In this sense, the global models discussed
below may serve as proxies for the modeling used for the entire
special section.

The remainder of the paper is organized as follows. In section 2,
we compare results from single fluid, plasma-only models, and in
section 3, from five neutral/plasma global heliospheric models for
one particular parameter set. In section 4, we attempt to qualify
our results and put them into perspective of the overall goal of
realistic models of the global heliosphere.

%__________________________________________________________________

\section{Comparison of plasma-only models}

The five self-consistent global heliospheric models that are
compared in this paper are the \citet{Baranov93} Moscow model
(``BM''), the IGPP-UCR kinetic model by \citet{Heerikhuisen06}
(``Hee'') and the multi-fluid model by \citet{Florinski05}
(``Flo'') extended from the \citet{Florinskietal03b} two-fluid
model, the \citet{Pauls95} style multi-fluid model modified by
\citet{Mueller06} (``Mue''), and the Bonn five-fluid model
\citep{Fahr00} as used by \citet{Scherer03} (``Sch''). All these
models use a gas-dynamic description of the plasma, and therefore
we start with a comparison of the plasma-only part, i.e.\ the
Euler equation solvers used by these groups for their plasma part
under the assumption that there are no source terms on the
right-hand-sides of the fluid equations (no neutrals in the
system). All groups ran their plasma code with the solar wind and
interstellar boundary conditions listed in Table 1. It was assumed
that the plasma consists of equal (comoving) densities of protons
and electrons, in other words, the thermal plasma pressure equals
twice the thermal proton pressure. The magnetic field, as well as
the solar gravity, were neglected.

%                                     Two column figure (place early!)
%______________________________________________ sing_1 (lg rho, lg e)
   \begin{figure*}
   \centering
   \includegraphics[width=\columnwidth]{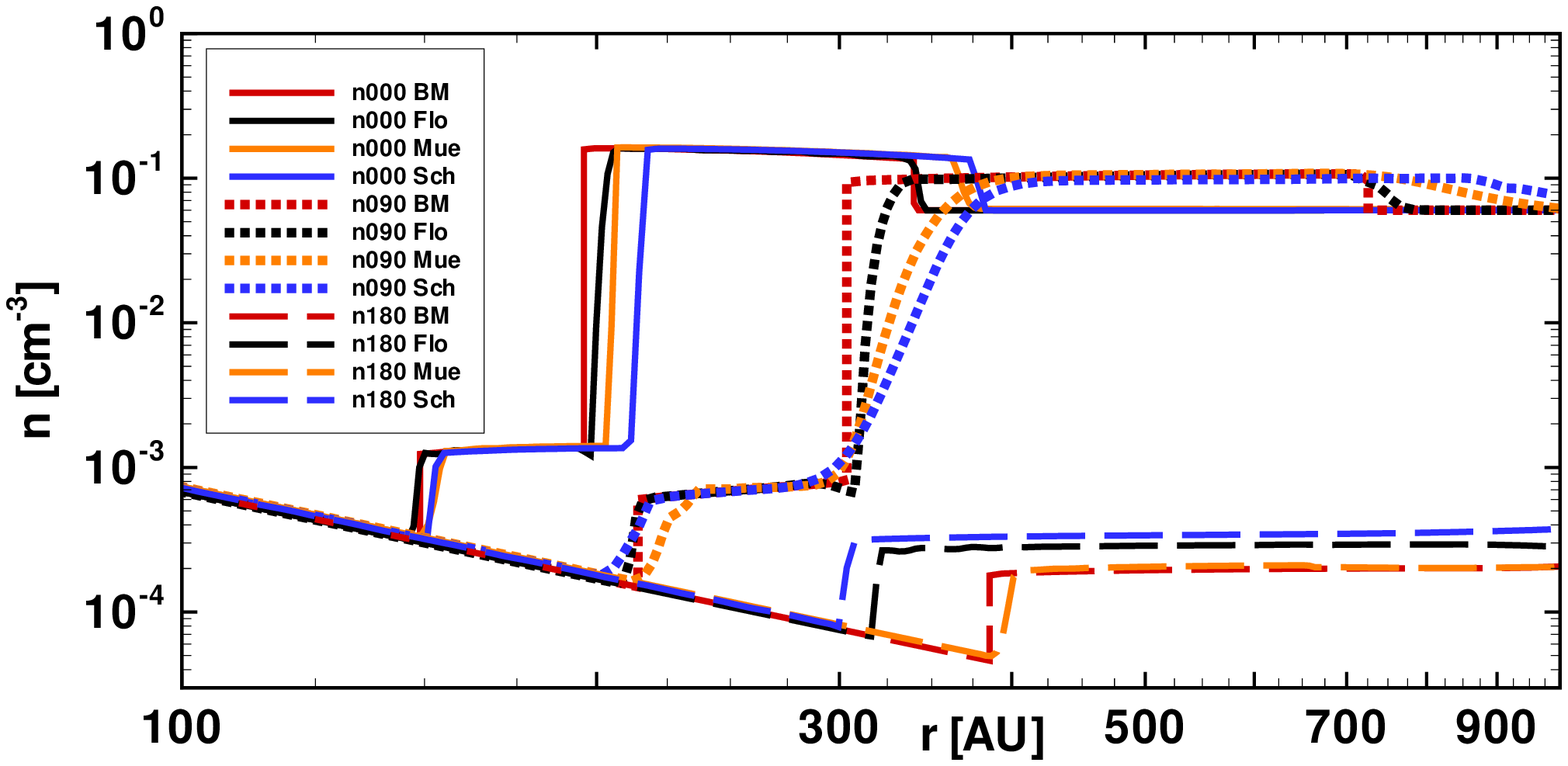}
   \includegraphics[width=\columnwidth]{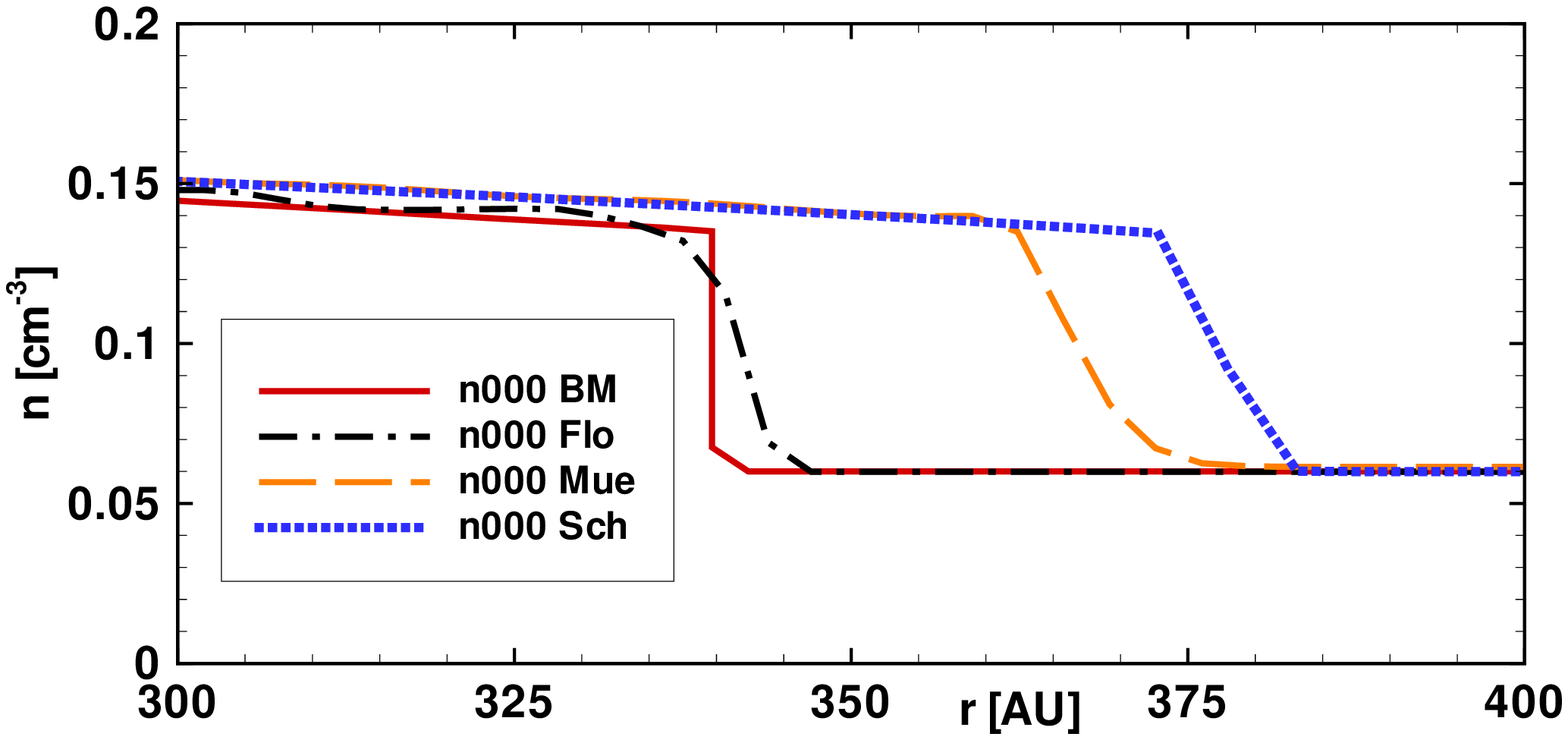}
   \caption{(a) Number density profiles for the plasma-only models of
   all four groups, in the directions upwind (0),
   crosswind (90), and downwind (180) with respect to the LIC flow, respectively.
   (b) Detail around upwind BS.}
              \label{figsnoverv}%
    \end{figure*}
%

%___________ simple table
   \begin{table}
      \caption[]{Boundary parameters, plasma-only models.}
         \label{bcplas}
     $$
         \begin{array}{p{0.3\linewidth}rrl}
            \hline
            \noalign{\smallskip}
            Variable      &  {\rm 1 AU} & {\rm LIC} & {\rm [units]} \\
            \noalign{\smallskip}
            \hline
            \noalign{\smallskip}
%            density & T / {[\mathrm{K}]} \leq 1700^{\mathrm{a}}     \\
            proton density & 7      &    0.06 & {[\mathrm{cm^{-3}}]}\\
            velocity     & 375    &   26.4  & {[\mathrm{km\, s^{-1}}]}\\
            temperature  & $73,640$ & 6530    & {[\mathrm{K}]} \\
            \noalign{\smallskip}
            \hline
         \end{array}
     $$
   \end{table}

As can be expected, the results from the four groups are very
close to each other (the plasma parts of the Hee and the Flo
models are identical, and not listed separately). Figure
\ref{figsnoverv} shows this with the number density profiles of
all four models, in three representative directions in the
commonly used heliocentric reference frame. In all three
directions, all the densities follow a $r^{-2}$ power law in the
supersonic solar wind before encountering the termination shock.
The upwind termination shock (TS) distances are very close to each
other, whereas downwind there is more variability. The shock
strengths (ratio of downstream to upstream density) are more or
less the same. Also the density contrast across the heliopause
(HP) is similar across the four models. Also obvious is that the
BM model uses capturing methods to identify and enforce
discontinuities, whereas the other three models do not employ such
techniques, and transitions are spread over a few grid points (see
the bow shock of the Mue model for an example, Fig.\
\ref{figsnoverv}b).

Table \ref{resplas} lists some key results for the shock and
heliopause locations. The similarities in the results are evident.
The last column comprises a simple average across the four models
for each result, and the standard deviation hints at the range
that the results span. The different models basically agree on the
upwind TS and HP locations, and are a little bit more spread for
the BS, and yet more for the downwind TS results.

%___________ table: key results 1
   \begin{table}
      \caption[]{Key results from plasma-only models.}
         \label{resplas}
     $$
         \begin{array}{p{0.3\linewidth}rrrrr}
            \hline
            \noalign{\smallskip}
            Result      & BM & Flo &  Mue & Sch & mean \\
            \noalign{\smallskip}
            \hline
            \noalign{\smallskip}
            upwind TS [AU] &  149  &  148 & 153  & 152  & 150.5\pm 2.4\\
            upwind HP [AU] &  196  &  199 & 204  & 212  & 202.8\pm 7.0\\
            upwind BS [AU] &  340  &  340 & 365  & 380  & 356\pm 18\ \\
            downwind TS    &  385  &  319 & 396  & 304  & 351\pm 46\ \\
            \noalign{\smallskip}
            \hline
         \end{array}
     $$
   \end{table}

Besides the treatment of shocks and discontinuities, there are
obviously many other reasons why the four models vary from each
other. Each of the four models makes different choices related to
the grid configuration, resolution, and the extent of the
simulation domain. Also, there are four different choices of the
numerical transport and diffusion schemes to solve the Euler
equations.
%Numerical Method
BM use a Godunov-type numerical scheme with moving adaptive grid
while capturing three discontinuities --- the heliopause as
contact discontinuity, and the termination and bow shocks. The
accuracy of the numerical scheme resolution is improved by using a
``minmod'' limiter. The plasma part of the numerical algorithm of
the Flo multi-fluid model uses the total variation diminishing
(TVD) finite volume scheme based on the Courant-Isaacson-Rees
approximate Riemann solver. Conservation laws for the neutral
components (section 3) are solved using the more diffusive TVD
Lax-Friedrichs method. The Hee plasma part is identical to that of
Flo. The ZEUS-3D algorithm underlying the Mue model is based on
the method of finite differences on a staggered mesh,
incorporating a van Leer monotonic advection scheme, and von
Neumann-Richtmyer artificial viscosity at shock fronts. For all
fluids in the Sch model the Euler equations are formulated for
quantities conserving the flux of mass, momentum, and energy, and
are subjected to second order Riemann solvers using the
Lax-Friedrichs method with an entropy fix. For large pressure
gradients a Harten-Lax-van Leer solver is implemented.

\begin{figure}
   \centering
   \includegraphics[width=\columnwidth]{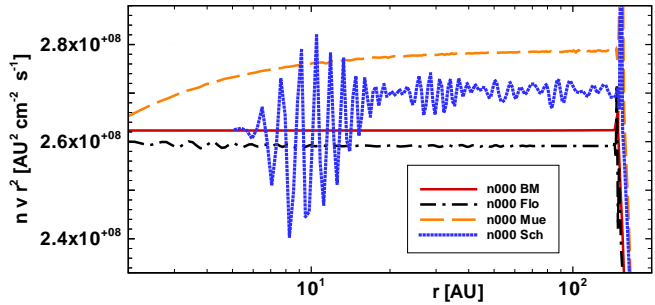}
\caption{Upwind profiles of flux conservation $n\, v\, r^2$ in the
supersonic solar wind.}
              \label{figsninner}%
    \end{figure}

The high-Mach number regime of the supersonic solar wind is an
instructive example of the modeling technique differences, and
their consequence for heliospheric studies. While each technique
is optimized to conserve crucial quantities (for example, mass
flux from grid cell to grid cell), the calculations of density,
velocity and pressure deviate from model to model. Small flux
errors are evident in Figure \ref{figsninner} showing the
conserved total particle flux $n v r^2$, where the ideal value
($2.625\times 10^8$ AU$^2$ cm$^{-2}$ s$^{-1}$) is approximated
well by the BM model. The Flo model also conserves this quantity,
albeit a smaller value was introduced at the boundary. Immediately
upstream of the TS, the modeled densities range from 6.75 to 7.35
cm$^{-3}/r^2$, and the velocities (ideally 375 km s$^{-1}$) range
from 376 to 383 km s$^{-1}$. As the location of the TS is
determined by the ram pressure at the TS balancing the ISM
pressure, the subtle variations in ram pressure are a natural
explanation of the TS differences in Table \ref{resplas}. Similar
effects explain the other discrepancies. The BS distances
basically follow the trend of the HP distances, as the BS shock
compression ratios are quite similar between all four models
(2.2--2.3; cf.\ Fig.\ \ref{figsnoverv}b).
%The location of the bow shock is quite
%sensitive to this shock strength, as post shock adiabatic
%deceleration still needs to achieve zero velocity at the
%stagnation point (HP).
We note in passing that the stagnation axis from which data for
Table \ref{resplas} and Figures \ref{figsnoverv} (except for
90$^o$) and \ref{figsninner} are taken, is numerically somewhat
problematic in that for the axisymmetric models at hand, it
actually consists of boundary grid zones and not of interior
zones.

%   fig2                          single column figure
%______________________________________________ sing_1 (lg rho, lg e)

One specific difficulty faced by every model of the heliosphere is
the treatment of outflow boundary conditions in the heliotail.
Because the tail plasma flow is subsonic, the boundary is
influenced by waves and disturbances entering from downstream,
i.e., from the regions not included in the simulation. Simple
outflow boundary conditions are used by the Mue and Sch models, in
effect copying interior solutions to boundary shadow zones and
thus making derivatives at the boundary small. This approach could
lead to waves reflected off the boundary and reentering the
simulation, an unphysical situation. A somewhat more complicated
approach, employed by the Flo model, is to apply a
``non-reflective'' outflow boundary condition, whereby the flow is
reaccelerated to a sonic point through an insertion of a
rarefaction fan at the boundary of the domain. In the BM model the
tail computation region is extended up to the region where the
solar wind is supersonic again. While the other models agreed to
an outer boundary at 1000 AU, the BM model extends the simulation
domain to 6000 AU tailward for this reason
\citep{Izmodenov03b,Alexashov04b}, at the expense of resolution
and computation time. Regardless of their degree of
sophistication, it should be realized that all tail boundary
conditions used by heliospheric models are not physically exact in
the strict sense, except for the BM model where the outer tail
flow is supersonic and, therefore, the boundary conditions are
correct. The boundary handling contributes to noticeable
differences in the distances of the TS in the downwind direction
predicted by different models (see Table \ref{resplas}), yet it is
not the only issue involved, as there is a curious pairing of BM
and Mue model distances on the the one hand, and Flo and Sch
models on the other hand.

%__________________________________________________________________

\section{Comparison of self-consistent plasma/neutral models}

\subsection{Model results}

We now proceed to introduce neutral interstellar hydrogen (H) into
the system and, using the plasma codes of section 2, switch on the
full, self-consistent plasma/neutral codes in which the plasma and
neutral H influence each other through appropriate source terms.
All groups calculate their global heliosphere with the solar wind
and interstellar boundary conditions listed in Table \ref{bcfull}.
Again, the magnetic fields are neglected, as are gravity and
radiation pressure. The H-p charge exchange cross section depends
on the relative velocity; for this paper all five models use the
\citet{MaherTinsley} cross section. A photoionization rate of
$10^{-8}$ s$^{-1}$ (1 AU / $r$)$^2$ is assumed, and other
ionization channels such as electron impact ionization are
neglected throughout.

%___________ simple table
   \begin{table}
      \caption[]{Boundary parameters, full models.}
         \label{bcfull}
     $$
         \begin{array}{p{0.3\linewidth}rrl}
            \hline
            \noalign{\smallskip}
            Variable      &  {\rm 1 AU} & {\rm LIC} & {\rm [units]} \\
            \noalign{\smallskip}
            \hline
            \noalign{\smallskip}
%            density & T / {[\mathrm{K}]} \leq 1700^{\mathrm{a}}     \\
            proton density      & 7      &    0.06 & {[\mathrm{cm^{-3}}]}\\
            H density      & -      &    0.18 & {[\mathrm{cm^{-3}}]}\\
            velocity     & 375    &   26.4  & {[\mathrm{km\, s^{-1}}]}\\
            temperature  & $73,640$ & 6530    & {[\mathrm{K}]} \\
            \noalign{\smallskip}
            \hline
         \end{array}
     $$
   \end{table}

%        fig 3                     Two column figure (place early!)
%______________________________________________ sing_1 (lg rho, lg e)
   \begin{figure*}
   \centering
   \includegraphics[width=\columnwidth]{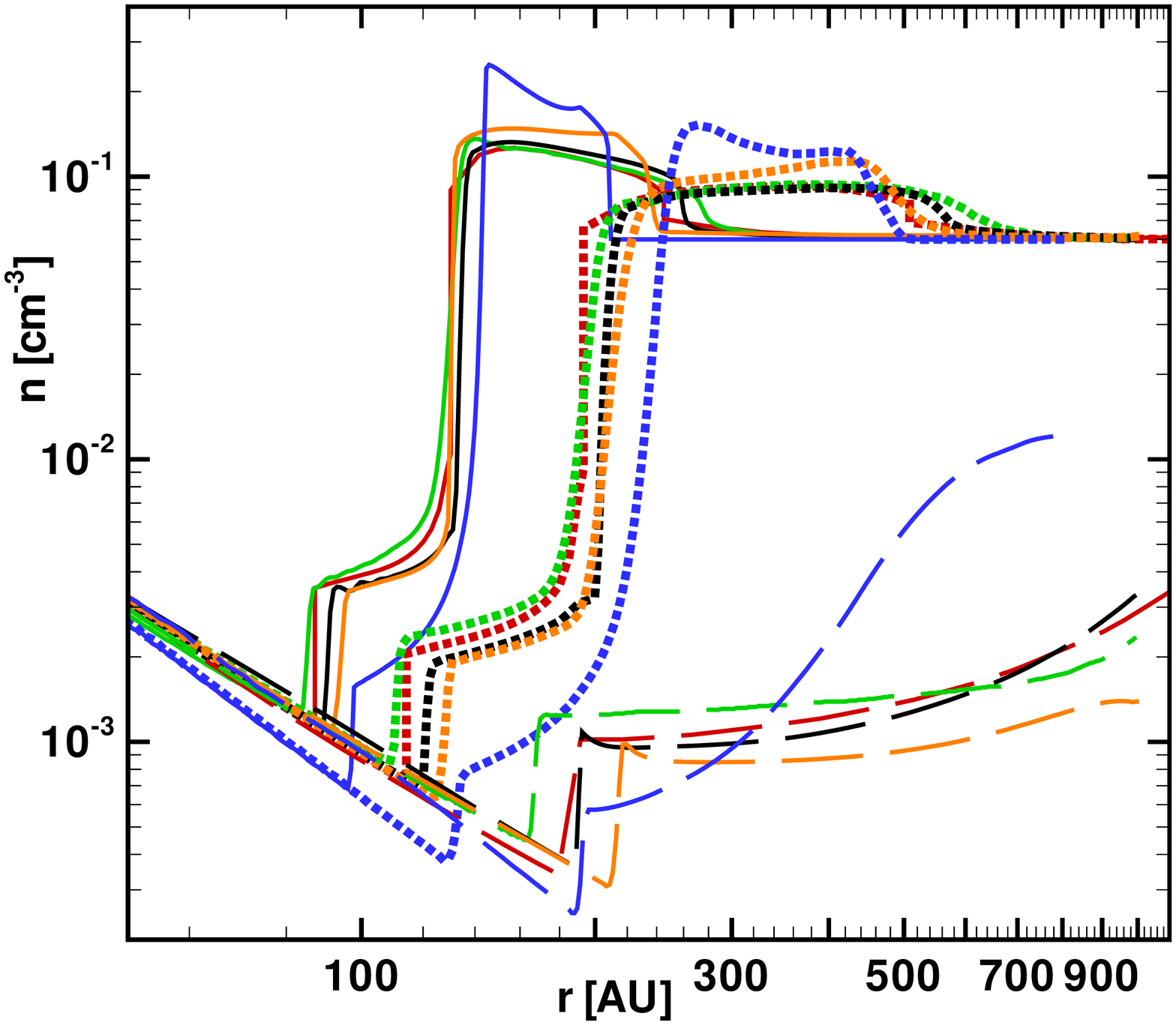}
   \includegraphics[width=\columnwidth]{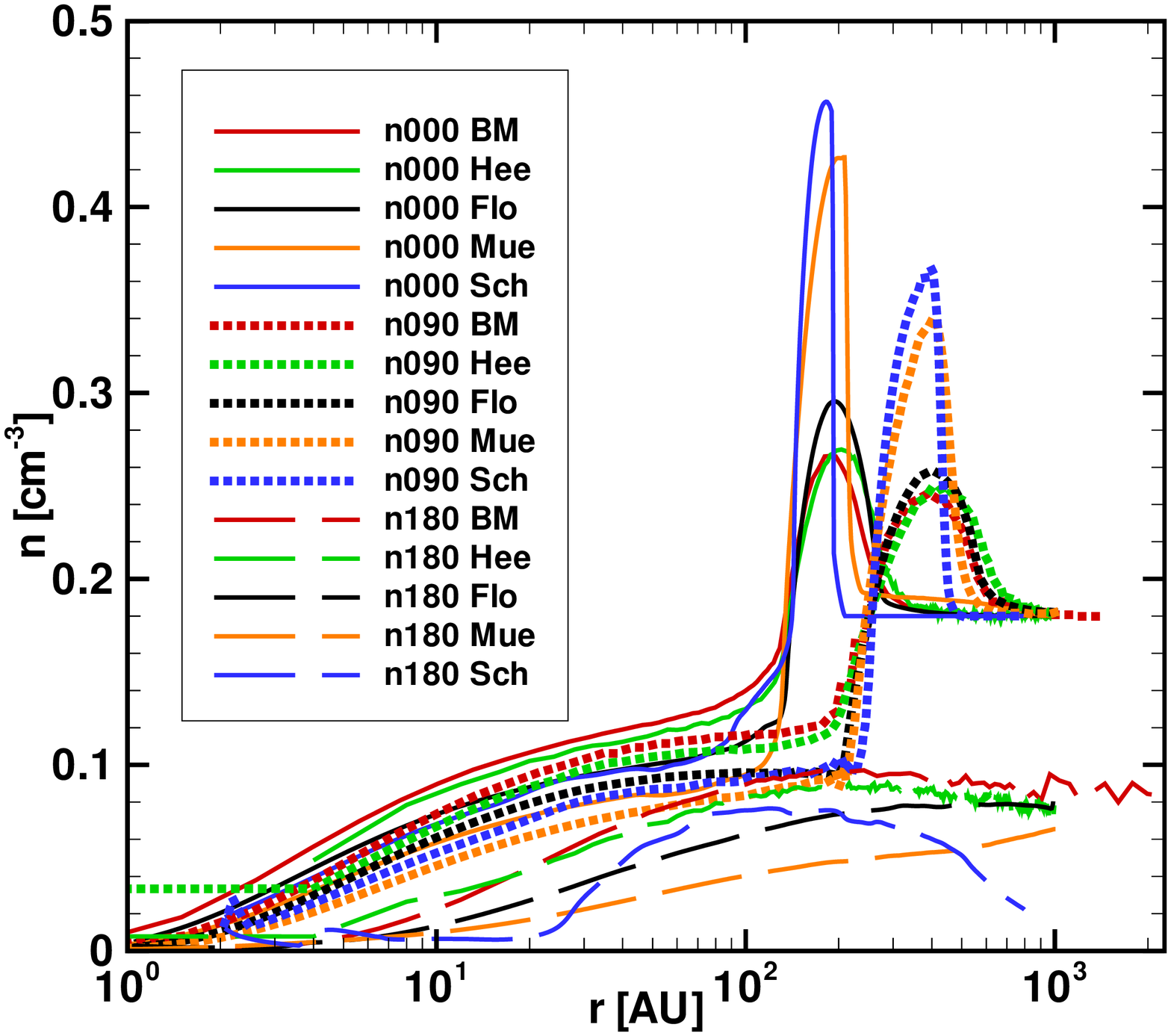}
\caption{Density profiles of plasma (left), and of neutral H
(right), for three directions, for the five self-consistent
plasma/neutral models.}
              \label{figninout}%
    \end{figure*}

Figure \ref{figninout} gives a good overview of the results for
the two available plasma - kinetic neutral models BM and Hee, and
the three multi-fluid models Flo, Mue, and Sch. The left panel
shows the plasma density profiles for upwind (solid), crosswind
(dotted), and downwind (dashed) directions, and the right panel
displays the information for total neutral density in the same
format. The plasma results exhibit a level of similarity to each
other that is comparable or slightly better than the level of
similarity of the plasma results in the previous section (Fig.\
\ref{figsnoverv}). In particular, the upwind HP location nearly
coincides in all five models. The first four entries of Table
\ref{resfull} contain the key locations of the heliosphere for the
full models. Again, the standard deviations in the last column
express the range of the results against a simple arithmetic mean
of the values in each row. It can be seen that also the upwind TS
locations agree to within 7\%, and only BS and downwind TS
disagree (up to 14\% each). While this type of disagreement was
also found in the plasma-only cases of section 2, it tends to now
be larger, especially for the bow shock.  We note in passing the
dramatic effect on the heliosphere boundary locations that the
inclusion of neutrals has. The results of Table \ref{resplas} are
significantly larger than those of Table \ref{resfull} even though
the boundary parameters of the plasma-only case are identical to
those of the plasma/neutral case.

%___________ table: key results 2
   \begin{table*}
      \caption[]{Key results from plasma/neutral models.}
         \label{resfull}
     $$
         \begin{array}{p{0.17\linewidth}rrrrrc}
            \hline
            \noalign{\smallskip}
            Result         & BM & Hee & Flo &  Mue & Sch & mean\\
            \noalign{\smallskip}
            \hline
            \noalign{\smallskip}
            upwind TS [AU] & 87 & 85 & 90  & 94  & 96  & 90.4\pm 4.6\\
            upwind HP [AU] & 130 &126 & 132  & 130  & 138  & 131.2\pm 4.4\ \\
            upwind BS [AU] & 245 & 274 & 260  & 236  & 209  & 245\pm 25\\
            downwind TS    & 177 & 166 & 190  & 214  & 192  & 188\pm 18\\
            BS compression ratio & 1.2 & 1.2 & 1.4  & 1.7  & 2.3  & 1.6 \pm 0.5\\
            peak $n_H$ [cm$^{-3}$] & 0.27 & 0.27 &  0.30  & 0.43  & 0.46 & 0.346\pm 0.092\\
            $n_H$ at TS [cm$^{-3}$] & 0.134 & 0.125 &  0.109  & 0.094  & 0.126 & 0.118\pm 0.016\\
%            $n_H$ at TS [cm$^{-3}$] & 0.13 & 0.13 &  0.11  & 0.09  & 0.13 & 0.12 \\
            filtration $f$ & 0.74 & 0.69 &  0.61  & 0.52  & 0.70 & 0.65\pm 0.08\\
            $v_H$ at TS [km s$^{-1}$]  & 20.7 & 20.8 & 23.4 & 21.3  & 19.2 & 21.1\pm 1.5\\
%            $T_H$ at HP [K]  & 25700 & 29300 &     & 24700  & 14200 & 20.5\\
            $T_H$ at TS [K]  & 26800 & 30900 & 15500 & 21000  & 12000 & 21200\pm 7800 \\
            \noalign{\smallskip}
            \hline
         \end{array}
     $$
   \end{table*}

In the neutral H density (Fig.\ \ref{figninout}, right) all models
exhibit an overdensity (hydrogen wall) downstream of the bow
shock, and a subsequent rapid drop in the density approaching the
heliopause and further inside. For this and similar diagnostics,
the neutral multi-fluid results are summed (averaged) into a
single total neutral hydrogen quantity, in the simplest manner as
$n_{tot} = \sum_i n_i$ for total density, $v_{tot} = n_{tot}^{-1}
\cdot \sum_i n_i v_i$ for velocity, and $T = n_{tot}^{-1} \cdot
\sum_i n_i T_i$ for temperature.

The two fluid models with less than four neutral fluids (Mue, Sch)
almost agree in the sharpness and the peak height of the hydrogen
wall. As in previous findings (\citet{Baranov98}; Figure 2 by
\citet{McNutt04}; AI05; HFZ06), the hydrogen wall is quite a bit
higher for these two fluid models compared to the kinetic models
BM and Hee. The two latter models match each other well in neutral
hydrogen. The hydrogen wall of five-fluid model Flo is higher than
the kinetic ones, but closer to those than to the other
multi-fluid models. The peak densities in the hydrogen wall are
listed in Table \ref{resfull}. The hydrogen wall profiles fit the
general trend displayed in Figure 3 of AI05: There, the one-fluid
model (most similar to the Sch model) resulted in the
highest-peaked hydrogen wall, the three-fluid model (most similar
to the Mue model) exhibited a somewhat smaller hydrogen wall with
a very sharp rise on the interstellar side, and the BM model had a
small peak, with a smooth H density rise and fall, that was
relatively closely matched by a four-fluid model (most similar to
the Flo model). Note that the neutral H column density through the
upwind direction is basically constant; the displayed different
hydrogen walls are either tall and narrow, or smaller and broad.

%          fig 4             Two column figure (place early!)
%______________________________________________ sing_1 (lg rho, lg e)
   \begin{figure*}
   \centering
   \includegraphics[width=\columnwidth]{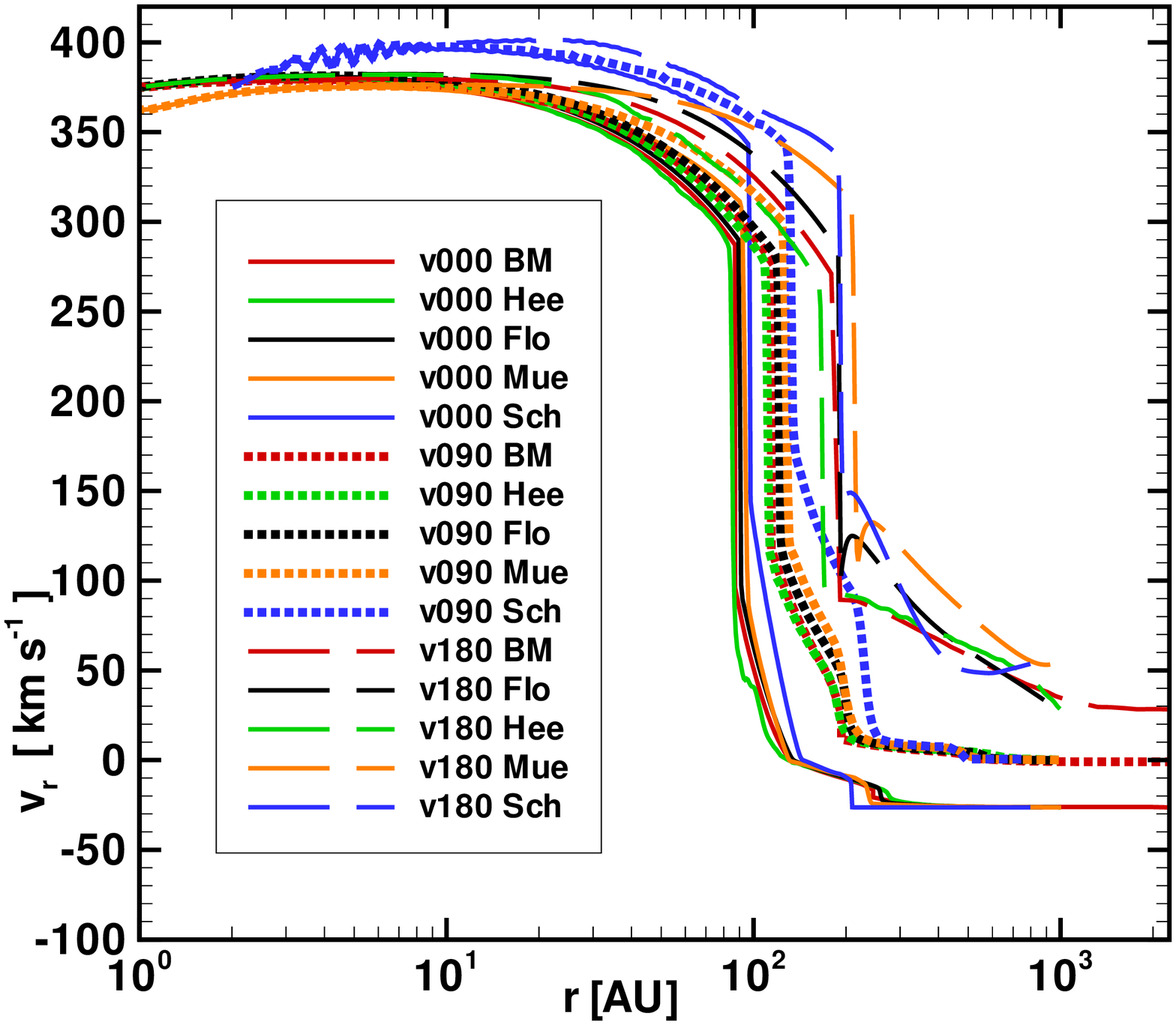}
   \includegraphics[width=\columnwidth]{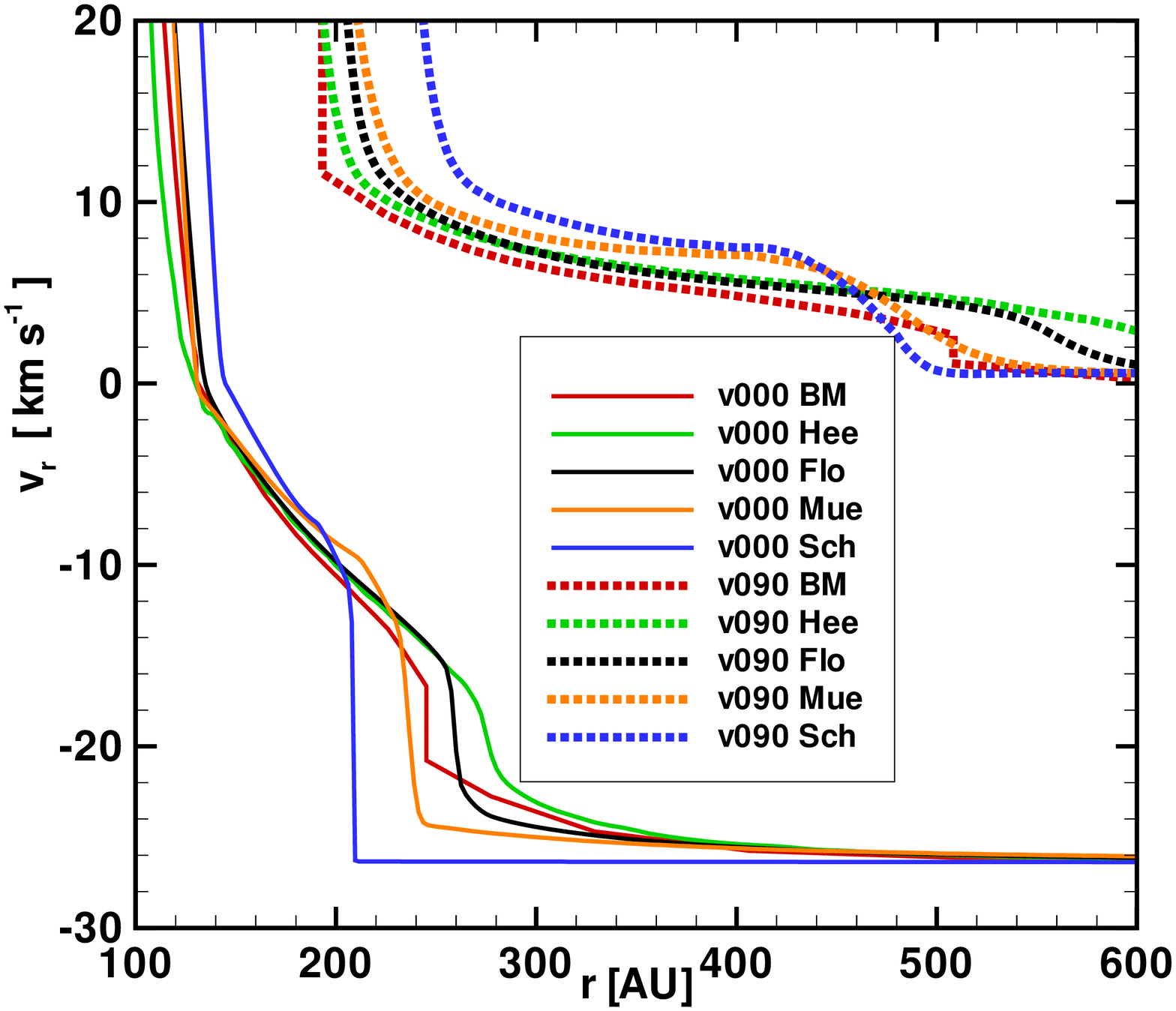}
\caption{Plasma radial velocity profiles of all models for three
directions (left), and a magnification around the upwind bow shock
(right).}
              \label{figvr}%
    \end{figure*}

The largest contributor to the differences between the simulated
hydrogen walls is the distribution of plasma velocities, notably
the component parallel to the ISM flow. Figure \ref{figvr} shows
an overview of the radial velocity component as a function of
heliocentric distance in the left panel, and the right panel zooms
in around the bow shock distance. The two kinetic models display a
plasma deceleration upstream of the bow shock due to charge
exchange with component 3 and 4 neutrals that are streaming
antisunward and have passed the bow shock. To a lesser extent, the
Flo and Mue models exhibit a similar deceleration, whereas the Sch
model cannot resolve such counterstreaming fluid elements (they
deposit their energy already far downstream of the BS). As a
consequence of the deceleration both upstream and downstream of
the BS, the BS is the weakest in the kinetic cases, followed in
shock strength by the Flo model, and is the strongest in the
one-fluid case, with the Mue model in between (Table
\ref{resfull}). The shock-capturing method of BM arrives at a very
weak BS. This range of bow shock strengths explains the more
gradual hydrogen wall in the kinetic cases. The hydrogen wall is
of lesser amplitude in the kinetic cases because the velocities
downstream of the BS are distinctly larger (absolute magnitude)
than those in the fluid cases, and therefore charge-exchanged
neutrals are not decelerated as much as in the one-fluid case
where the plasma velocity is the smallest. To appreciate this
difference, one has to mentally shift the plots of Fig.\
\ref{figvr} (right) so that the individual bow shocks line up. As
expected, a stronger bow shock results in a more decelerated
plasma, and therefore a larger peak density of the hydrogen wall.
Typically, this also means a lesser distance of the BS to the Sun,
and this trend is evident in Table \ref{resfull}. The exception is
model BM that experiences additional deceleration downstream of
the BS, such that the BS standoff distance is not as far outward
as the shock strength would suggest.

The different hydrogen walls result in different neutral densities
downwind of the TS, where the neutrals enter region 4 of the
supersonic solar wind. The filtration ratio $n_H(TS)/n_H(\infty)$
is listed in Table \ref{resfull}, along with the two absolute
densities discussed above. In principle, it can be expected that
the slower the outer heliosphere plasma is (i.e. the higher the
hydrogen wall), the more neutrals get deflected around region 4
and hence the stronger the filtration is. However, for the slowest
case, the one-fluid Sch model, the small width of the inner
heliosheath compensates for this effect and creates TS neutral
densities similar to the kinetic BM model.
\citet{Richardson08} investigate the slowdown of the supersonic
solar wind due to the pickup process derived from neutrals. They
calculate a 15\% solar wind slowdown between 5 AU and 78 AU from a
2006 conjunction of the Ulysses and Voyager 2 spacecraft. The five
models described here indicate less of a slowdown in this distance
range, namely, between 7\% (Mue, Sch) and 10\% (all others).

While the differences in the neutral density at the termination
shock (and consequently, the differences in the filtration ratio)
are in the 20\% range, the predicted neutral velocity at the
termination shock varies much less among the five models. Table
\ref{resfull} contains the corresponding results, as well as those
referring to the (total) neutral temperature at the termination
shock. The velocities are within 6\% of each other; in other
words, all five models predict a slowdown of the neutrals by
($5.3\pm 1.5$) km s$^{-1}$ because of their passage from the LIC
to region 4. The picture is much less clear for the neutral
temperature, where large variations exist between the two-fluid
model (smallest $T$) on the one side, and the kinetic models
(largest kinetic $T$) on the other side, with the Mue model in
between.

\subsection{Discussion}

As is evident from the results above, the two kinetic models yield
very similar results, while the discrepancies between the kinetic
models and the one-fluid model are the largest, which was also
found by AI05. These two modeling alternatives bracket the range
within which the rest of the models fall.  The kinetic codes have
no provision for direct neutral--neutral interaction, and hence
have no direct, built-in mechanism that would drive the neutral
distributions toward Maxwellian equilibrium.  On the other hand,
the Euler equations of gas dynamics are derived under the
assumption of frequent particle collisions (even though they seem
to be valid beyond that range) to yield thermodynamic equilibrium,
and hence the single-fluid description of neutral H is related to
envisioning frequent neutral collisions. The multi-fluid codes are
interesting in that they disallow neutral--neutral interactions
between neutrals that are created in different thermodynamic
regimes (which usually means that they are distinctly separated in
velocity space), while still envisioning neutral collisions within
those regimes themselves.
As neutral H is being modeled with more and more fluids, fewer and
fewer neutral collisions are assumed, which results in a
convergence toward the kinetic results, seen in the results above
and those by AI05 and HFZ06.

The absence of neutral--neutral interactions, in particular in the
kinetic models, does however not necessarily mean that the
distributions are completely non-Maxwellian. Charge exchange with
plasma protons injects neutrals drawn from a Maxwellian
distribution, hence the secondary neutrals have a tendency to
organize in distributions similar to Maxwellians. In this sense,
charge exchange constitutes an indirect (and inefficient) channel
for neutral equilibration.
The charge exchange mean free path (mfp) is nowhere very small in
the heliosphere, but is sometimes small enough for charge exchange
to occur frequently, driving neutrals toward equilibrium. Indeed,
some example mfps derived from the Mue model are quite short. In
the region upstream of the bow shock, the interstellar neutral mfp
is $\sim 200$ AU, and the mfp of neutrals having been generated in
regions 3 and 4 (the regions occupied by solar wind) and having
streamed to upstream of the bow shock is even smaller, $\sim 100$
AU. For the outer heliosheath (between BS and HP), mfps are below
100 AU, and go down even to 20 AU close to the nose of the
heliopause (on the interstellar side) where plasma velocities
become small. Therefore, the Knudsen number in these regions is
small, down to $\sim 0.2$. In these instances, and even when
Knudsen numbers are larger, forcing the neutrals into
multi-component fluids with Maxwellian distributions in general
does not change their distribution much. This interpretation is
backed up by the findings by HFZ06 who decompose the neutral
distribution function into the contributions from the four
components. They find that the interstellar component (component
1) and the outer sheath component 2 do behave like Maxwellians
even when treated fully kinetically (cf.\ their figure 5). In this
context, AI05 argue that one of the more fundamental differences
between kinetic and fluid treatments of neutral H is related to
the fact that for the interstellar temperature of 6530 K, the
thermal velocity of H is already about half of the bulk velocity
value of 26 km/s. This means that individual ISM particle
trajectories have sizeable perpendicular velocity components that
are represented in the kinetically modeled trajectories. In
contrast, the fluid description of the same region has a strictly
parallel bulk velocity, and effects of the perpendicular particle
motion are handled by a non-zero neutral pressure.

For secondary neutrals produced in the solar wind (regions 3 and
4), the fluid picture is capturing some aspects of the particle
behavior less well. Component 3 might be reasonably approximated
as a hot Maxwellian in region 3, but component 3 streaming out of
this region will have a complicated distribution function. For
points outside the HP, the distribution function will be
resembling a half-Maxwellian, with velocity components toward the
HP missing (half-Maxwellian in region 1, HFZ06). For region 4,
where component 3 neutrals constitute the energetic neutral atom
(ENA) hydrogen background, the distribution is complex (each
location is reached in principle by ENA from all heliosheath and
heliotail positions), and certainly non-Maxwellian. The fluid
approach consequently has component 3 very hot and with a small
velocity in region 4. Similar findings apply to component 4
neutrals, which are cold and fast in principle. As they stream to
distant locations in region 1, the kinetic codes allow their
distribution function to broaden unhindered by interactions and
thereby gaining a large kinetic temperature, whereas the fluid
component 4 experiences the adiabatic cooling of the radial
expansion, and ends up with much smaller temperatures. The
differences in component 3 and 4 neutrals present in region 1
between the kinetic and the fluid picture set the stage for the
different bow shock strengths mentioned above and hence influence
the BS location and the hydrogen wall. In the solar wind region,
the absolute energy transfer to the plasma due to charge exchange
by component 3 and 4 neutrals seems insensitive to the subtle
differences in the distribution function there, and hence TS and
HP locations are basically unaffected.

%__________________________________________________________________

\section{Sources of Error}

Using a multi-fluid approach instead of a particle kinetic method
incurs a systematic error in the neutral distributions, and
therefore also in the plasma distributions. This has been
discussed in the previous section and in the literature (e.g.,
AI05 and HFZ06).  In this section we want to discuss the source of
other systematic errors that contribute as well to differences
between any modeled global heliosphere and the real system as
observed through heliospheric measurements.

\subsection{Numerics}

As illustrated in section 2, simple choices for the fundamental
algorithm for following the non-MHD fluid equations, combined with
choices for grid resolution and organization (e.g., spherical vs.\
Cartesian, or fixed resolution vs. location-dependent) and choices
relating to the extent of the computation domain determine the
outcome of even the simplest, plasma-only heliospheric simulation.
Different choices will conserve different quantities better,
usually at the expense of other quantities (see the conservation
of $n v r^2$ in Fig.\ \ref{figsninner}).

Another common issue of fluid simulations is the handling of
discontinuities such as termination shock, bow shock, and
heliopause. The solutions used by the models in section 2 range
from smearing out discontinuities over three grid cells to shock
capturing methods that supply the discontinuous solution
externally, and not from the fluid algorithm used everywhere else.
The treatment of discontinuities is hence quite sensitive to the
local grid resolution at the heliospheric boundaries.

As is usually the case, both multi-fluid simulations as well as
billion-particle simulations involve a myriad of individual steps,
with the potential that even numerical accuracy comes into play as
a potential source of error. This is presumably less important,
however, as the simulations eventually settle into a converged,
time-independent state for which roundoff errors should cancel.

\subsection{Cross sections}

The results of global heliosphere modeling are sensitive to the
cross sections that are chosen for the resonant charge exchange
between protons and neutral hydrogen. Often, studies of this
charge exchange cross section have been motivated by
investigations of the terrestrial ionosphere interacting with the
terrestrial neutral exosphere, and hence are not meant for higher
energies $> 1$ keV, which is a source of error for charge exchange
involving component 3 and 4 neutrals.

\citet{Izmodenov00} and \citet{Fahr07} have reviewed issues
relating to the relevant cross sections. Typically heliospheric
modelers have adhered to the energy-dependent cross sections by
\citet{MaherTinsley} and \citet{Fite}. Both are fitting formulae
of the form $(a - b \log v)^2$, where $v$ is the relative velocity
between the interaction partners. A recent compilation by
\citet{lindsey05} arrives at a yet different cross section
approximation. The cross sections are still uncertain to
approximately 10\% (solar wind speeds) and up to 40\% (slow
speeds; see, e.g., \citet{Bzowski08}), not only because of the
fitting itself and the extrapolation of these fits beyond their
intended velocity range, but also the underlying experimental data
from different groups do not always reconcile easily.

Heliospheric modeling is very sensitive to the actual cross
section values. In order to not repeat work reported elsewhere, we
would like to draw attention to figure 4 by HFZ06 and figure 8 by
\citet{Baranov98}. Each of these two figures compares two
(respective) heliospheric models that differ only by the choice of
the cross section, i.e.\ either using the values by
\citet{MaherTinsley} or those by \citet{Fite}. For both papers,
the results indicate that the shift in heliospheric boundary
locations like TS and HP is minor, about 1--3\%, but that the
consequences for the hydrogen wall and for the neutral filtration
factor are quite a bit larger: The hydrogen wall is $\sim 14$\%
larger in the \citet{Fite} case, and correspondingly, there is
more filtration going on for that case (smaller $f$ number, as
used in Table \ref{resfull}). The reason is that the \citet{Fite}
cross section is larger than the \citet{MaherTinsley} cross
section at key energies.

In this sense, the uncertainty in the cross section and issues
related to it are one of the larger error sources influencing
global modeling and its comparison to direct measurements. Note
that many data products derived from direct measurements use a
charge exchange cross section as well, as part of the ionization
channels acting on neutrals. Therefore, the discussion of this
systematic error applies to these derived data products as well.
For examples see \citet{Bzowski08} and \citet{Richardson08} in
this issue, where the uncertainty about the charge exchange cross
sections is echoed in the interpretation of the pickup ion
results, or the solar wind slowdown results, respectively, in
terms of the interstellar H density.

\subsection{Additional physics}

Finally, there are systematic errors influencing global
heliospheric modeling whose magnitude is difficult to assess or
sometimes not yet explored. Neglecting interplanetary and
interstellar magnetic fields, for example, excludes a whole suite
of possible heliospheric asymmetries, shifts in the heliospheric
boundaries, and influences on the neutral hydrogen distribution
even in the inner heliosphere. The presence of magnetic fields
also typically allows for temperature anisotropies and turbulence
in the plasma, and there is evidence that the solar wind (plasma)
velocity distribution is non-Maxwellian already at a 1 AU
distance, which most global models do not yet address. Further
away from the Sun, the proton distribution functions are driven
away from equilibrium by the effects of charge exchange, which
calls for a fully kinetic plasma -- neutral gas numerical modeling
strategy eventually.

The 3D, time-dependent solar wind in real-time differs from what
most models currently feed into their simulations. Similarly, the
solar irradiance depends non-trivially on time and on position in
the heliosphere. Additional simplifying assumptions often made
include the restriction of the particle species to electrons,
protons, and neutral hydrogen, and omitting heavier ion species
(including alpha particles) in the solar wind, and heavier
particles in the interstellar medium.
The influence of high-energy particles such as cosmic rays
(anomalous and galactic) should be taken into account, however
their effect on the heliospheric structure is most likely not
significant.

The pickup process, i.e., the dynamical process of accelerating a
newly born ion into the plasma bulk flow, also is often not
handled in sufficient detail in global models. Many times, global
modeling assumes instantaneous pickup for simplicity. It would be
more realistic to fairly treat the pickup ion evolution,
accounting for plasma-wave or turbulent interaction, and in
general accounting for the non-Maxwellian character of the pickup
ion distribution. A first level of refinement is taken in the Sch
model \citep{Fahr00} as used in this paper (section 3), where
pickup ions are not absorbed instantaneously into the main plasma,
but followed as a separate plasma fluid which interacts with the
main solar wind protons.

This list is not comprehensive, but is meant as a sample of the
type of issues that are outstanding for the business of global
heliospheric modeling, nonwithstanding past and present progress
on multiple fronts (numerous citations are omitted here for the
sake of brevity). Further progress, as well as extensions and
refinements of additional lines of model physics, will improve the
realism of all the models over time.

%__________________________________________________________________

\section{Conclusions}

We investigate in this paper global heliospheric plasma/neutral
models from five groups, first the plasma parts by themselves,
then the fully self-consistent models. For the latter, the neutral
species are modeled with either one, three, or four fluids, or on
a particle-kinetic level. Performing model runs with exactly the
same boundary parameters and physics included, we arrive at the
following conclusions.

\begin{enumerate}

\item Although very different numerical strategies and
approximations have been chosen for the five heliosphere models
presented in this paper, the results all qualitatively agree. In
many respects, even the models' quantitative predictions are
similar. They agree in particular about the location of the upwind
termination shock and the upwind heliopause. The discrepancies for
termination shock and heliopause in the five investigated models
range from a few percent in the nose direction ($<$7\%) to $<$14\%
in the downwind direction. The upwind distance of the bow shock
disagrees by up to 15\%. Also largely independent of the modeling
strategy in this sense is the velocity of neutrals entering region
4 through the upwind termination shock ($<$11\%).

\item The pileup of neutral H in the hydrogen wall is sensitive to
the modeling strategy, and the maximum density of neutral H
differs by about 60\% between the extreme cases. The column
density through the hydrogen wall does not seem to vary; the
hydrogen wall is either steep and narrow or small and broad. The
strength of the interstellar bow shock and the associated
post-shock velocity is the driver for the height of the hydrogen
wall, and the same mechanism leads to a variation in the
filtration, with the smaller hydrogen wall generally leading to
less filtration (larger neutral density entering through the
termination shock). The neutral H distribution in the inner
heliosphere is therefore moderately sensitive to the strength of
the interstellar bow shock. This is remarkable as the bow shock is
at the farthest heliospheric distance.

\item The strength of the bow shock also anticorrelates with its
resulting distance from the sun. In comparing the five modeling
strategies, the bow shock strengths differ by 90\% between the
extremes of the five models. The bow shock is strongest for a
two-fluid model, and turns out progressively weaker if neutrals
are modeled with more and more fluids, and is weakest in the
kinetic models. This behavior influences the neutral results in a
systematic way, with the filtration being the strongest in the
four-fluid case, weaker for the five-fluid model, and weakest in
the kinetic models. There are exceptions, however, as the
two-fluid model presented here yields a filtration as weak as the
kinetic models.

\item There are two discernible reasons for the different bow
shock strengths in the models. First, different numerical
strategies are used to model the bow shock itself, ranging from
shock-capture methods to smeared-out discontinuities. Second,
depending on the neutral modeling strategy, different amounts of
secondary neutral hydrogen make it to the region upstream of the
bow shock, and these neutrals have a much larger kinetic
temperature when modeled with particle-kinetic methods rather than
fluids. Both these factors prime the interstellar plasma through
charge exchange upstream of the bow shock, and weaken the bow
shock.

\item Global heliospheric models without neutrals (section 2) do
not reproduce many of the salient plasma density, velocity, and
temperature features of the heliosphere evidenced by models with
self-consistent neutrals (section 3), as can be seen, for example,
by comparing Figure \ref{figsnoverv}a to Figure \ref{figninout}a.
Naturally, the absence of the contribution of ISM neutral ram
pressure to the pressure balance leads to an enlarged heliosphere
in the plasma-only case (Table \ref{resplas} vs. Table
\ref{resfull}). This hence underlines the fact that the inclusion
of neutrals in a global heliosphere model -- in any
self-consistent way -- is critical to achieving physically
meaningful results.

\end{enumerate}

The fluid and kinetic neutral atom models agree to within about
10\% in some quantifiable measures, such as the location of the
principal heliospheric discontinuities, which is similar to the
uncertainties due to numerical algorithms, cross sections, grid
size and resolution. Larger differences of about 50\% exist in the
details of hydrogen distribution function (hydrogen wall magnitude
and neutral velocity distribution functions) between the models
based on kinetic and hydrodynamic neutral atom descriptions. The
uncertainties in our knowledge of the interstellar conditions, of
charge exchange cross sections, inclusion of the MHD effects
missing from the models discussed, and the unexplored effects of
additional physics are further expected to modify the results by a
similar amount.

%
%______________________________________________________________

\begin{acknowledgements}
The authors thank the International Space Science Institute
(ISSI), Bern, Switzerland, for enabling this study by hosting and
partially funding an International Team entitled ``Determination
of the Physical Hydrogen Parameters of the LIC from within the
Heliosphere.''  The authors also want to thank their respective
colleagues that have been instrumental in development of the codes
and perspective on heliospheric physics: V.\ Baranov, T.\ Kausch,
Y.\ Malama, L.\ Pauls, N.\ Pogorelov, and G.\ Zank. HRM
acknowledges partial funding through NASA SHP grants NAG5-12879
and NNG06GD55G. VF was supported in part by NASA grant NNG06GD43G,
and JH and HRM by NASA grant NNG06GD48G. VI and DA were supported
in part by RFBR grants 05-02-22000, 07-02-01101, 07-01-00291 and
by Program of Fundamental Research of OEMMPU RAS. V.I. also
acknowledges a partial support by the ``Dynasty'' Foundation. K.S.
acknowledges support by the Deutsche Forschungsgemeinschaft (DFG)
through project FI 706/6-2 ``Heliocauses,'' carried out within the
framework of the DFG priority program 1176 within CAWSES.
\end{acknowledgements}

%__________________________________________________________________

%\bibliographystyle{aa} % style aa.bst
%\bibliography{hrm_refs} % your references Yourfile.bib

\hyphenation{Post-Script Sprin-ger}

\end{document}